\documentclass[10pt,aps,prb,reprint,superscriptaddress,floatfix]{revtex4-1}

\usepackage[UKenglish]{babel}
\usepackage[utf8]{inputenc}
\usepackage[T1]{fontenc}

\usepackage{graphicx}
\usepackage{amssymb}
\usepackage{amsmath}
\usepackage{hyperref}
\usepackage{enumitem}
\usepackage{color}
\usepackage{nicefrac}
\usepackage{epstopdf}
\usepackage[amssymb]{SIunits}
\newif\ifuseexttikz
\useexttikztrue

\ifuseexttikz
\else
\usepackage{pgfplots}
\usepackage{tikz}
\usetikzlibrary{external} 
\usetikzlibrary{shapes}
\usetikzlibrary{calc}
\usetikzlibrary{external}
\fi

\hypersetup{pdfauthor={C. Hubig, J. Haegeman, U. Schollwoeck},pdftitle={Error estimates for extrapolations with matrix-product states}}

\newcommand{\tikzpic}[2]{
  \ifuseexttikz
  \includegraphics{#1}
  \else
  \tikzsetnextfilename{#1}
  #2
  \fi
}

\begin{document}
\title{Error estimates for extrapolations with matrix-product states}
\author{C.~Hubig}
\email{c.hubig@physik.uni-muenchen.de}
\affiliation{Department of Physics and Arnold Sommerfeld Center for Theoretical Physics, Ludwig-Maximilians-Universit\"at M\"unchen, Theresienstrasse 37, 80333 M\"unchen, Germany}
\author{J.~Haegeman}
\affiliation{Department of Physics and Astronomy, Ghent University, Krijgslaan 281 S9, B-9000 Ghent, Belgium}
\author{U.~Schollw\"ock}
\affiliation{Department of Physics and Arnold Sommerfeld Center for Theoretical Physics, Ludwig-Maximilians-Universit\"at M\"unchen, Theresienstrasse 37, 80333 M\"unchen, Germany}
\begin{abstract}
  We introduce a new error measure for matrix-product states without
  requiring the relatively costly two-site density matrix
  renormalization group (2DMRG). This error measure is based on an
  approximation of the full variance
  $\langle \psi | ( \hat H - E )^2 |\psi \rangle$. When applied to a
  series of matrix-product states at different bond dimensions
  obtained from a single-site density matrix renormalization group
  (1DMRG) calculation, it allows for the extrapolation of observables
  towards the zero-error case representing the exact ground state of
  the system. The calculation of the error measure is split into a
  sequential part of cost equivalent to two calculations of
  $\langle \psi | \hat H | \psi \rangle$ and a trivially parallelized
  part scaling like a single operator application in 2DMRG. The
  reliability of the new error measure is demonstrated at four
  examples: the $L=30, S=\nicefrac{1}{2}$ Heisenberg chain, the $L=50$
  Hubbard chain, an electronic model with long-range Coulomb-like
  interactions and the Hubbard model on a cylinder of size
  $10 \times 4$. Extrapolation in the new error measure is shown to be
  on-par with extrapolation in the 2DMRG truncation error or the full
  variance $\langle \psi | ( \hat H - E )^2 |\psi \rangle$ at a
  fraction of the computational effort.
\end{abstract}
\date{\today}
\maketitle
\section{\label{sec:intro}Introduction}

The density matrix renormalization group\cite{white92:_densit,
  schollwoeck11} (DMRG) method and its underlying matrix-product state
(MPS) structure are the method of choice for ground-state search and
representation of one-dimensional quantum states. In the last few
years, it has also been applied to wider and wider cylindrical
systems\cite{white04:_check_j, white09:_pairin_j,
  depenbrock12:_natur_spin_liquid_groun_state,
  yan11:_spin_liquid_groun_state_s, ehlers17:_hybrid_hubbar,
  motruk16:_densit} to mimic two-dimensional physics. Furthermore,
methods relying on the precise solution of a small effective system,
such as the dynamical mean-field theory\cite{georges92:_hubbar,
  karski08:_singl_mott_hubbar, ganahl15:_effic_dmft,
  wolf15:_imagin_time_matrix_produc_state,
  bauernfeind17:_fork_tensor_produc_states} (with or without a
dynamical cluster approximation) or the density matrix embedding
theory\cite{knizia13:_densit_matrix_embed} have also started to use
DMRG to solve the effective problem resulting from the
embedding. Growing computational resources as well as algorithmic
improvements made the study of critical systems\cite{white96:_dimer,
  yamashita98:_su, andersson99:_densit,
  vidal03:_entan_quant_critic_phenom, vidal07:_entan_renor,
  chen15:_quant_critic_spin_chain_emerg_su_symmet,
  khait17:_doped_kondo_luttin} in one dimension also more feasible.

In those complex systems it is often not possible to increase the
precision of the matrix-product state ansatz sufficiently to capture
the ground state of the system exactly. Instead, one often measures
both the observables of interest, among them the energy, and the
truncation error as obtained from a two-site DMRG (2DMRG) calculation
\emph{during} the calculation and at various precisions. One may then
extrapolate\cite{legeza96:_accur, white05:_densit,
  leblanc15:_solut_two_dimen_hubbar_model, ehlers17:_hybrid_hubbar}
the measured observables towards zero truncation error to obtain a
comparably accurate estimate of the ground-state observable. This
truncation error can also be obtained from a traditional
environment-site-site-environment DMRG procedure which is equivalent to the
MPS-based 2DMRG method.

Unfortunately, the 2DMRG method is relatively computationally
expensive,\cite{hubig15:_stric_dmrg} scales relatively badly in the
local physical dimension\cite{dorfner17:_numer} and it is sometimes
slow to pick up long-range correlations.\cite{white05:_densit} It
would hence be preferable to only use single-site DMRG (1DMRG,
corresponding to a environment-site-environment setup in the
traditional DMRG) for an approximately four-fold computational
speed-up in spin and fermionic systems and a much larger speed-up in bosonic
systems. The subspace expansion scheme\cite{hubig15:_stric_dmrg} for
1DMRG does not yield a usable truncation error. The related density
matrix perturbation\cite{white05:_densit} again scales relatively
badly in the local physical dimension.

Measuring the full variance
$\langle \psi | (\hat H-E)^2 | \psi \rangle$ would provide a reliable
error measurement for 1DMRG but is computationally very costly and
often impossible to evaluate even if expectation value measurements
and single-site DMRG calculations are still feasible. This is in
particular true for large systems with long-range interactions or an
underlying two-dimensional structure. Since such systems typically
result in highly entangled ground states and hence require large
computational resources per se, minimization of these resources
wherever possible is key.

For these reasons, we wish to formulate a method which measures an
error quantity $\mathit{err}(|\psi\rangle, \hat H)$ based only on a
matrix-product state $|\psi\rangle$ (regardless of how it was
obtained) and associated Hamiltonian $\hat H$. Measuring this error as
well as an observable for different states should allow an
extrapolation of the observable towards zero error. Evaluating the
error measure should not be much more costly than a 1DMRG
calculation. We find that the \emph{2-site variance}, an approximation
of the full variance, fulfills these requirements and turns 1DMRG into
a fast method with a controlled extrapolation scheme even for complex
systems.

The rest of the paper is structured as follows: In
Sec.~\ref{sec:notation} we briefly review MPS and the related
matrix-product operator (MPO) notation. Sec.~\ref{sec:alternatives}
discusses the currently available error measures. Sec.~\ref{sec:prop}
explains the new approximation of the variance to serve as the new
error measure. In Sec.~\ref{sec:examples}, we consider four relevant
examples to show that the variance itself is an error measure suitable
for extrapolations, that the 2-site approximation of the variance is
also a valid extrapolation tool even if it does not coincide with the
variance and finally that the 2-site variance is also applicable in
two-dimensional systems where evaluation of the full variance is not
possible any more. The conclusions in Sec.~\ref{sec:conclusions} serve
as a brief summary.

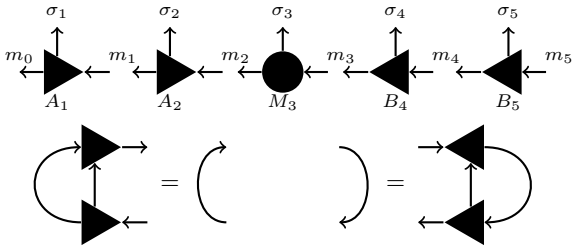
\begin{figure}[b]
  \centering
  \tikzpic{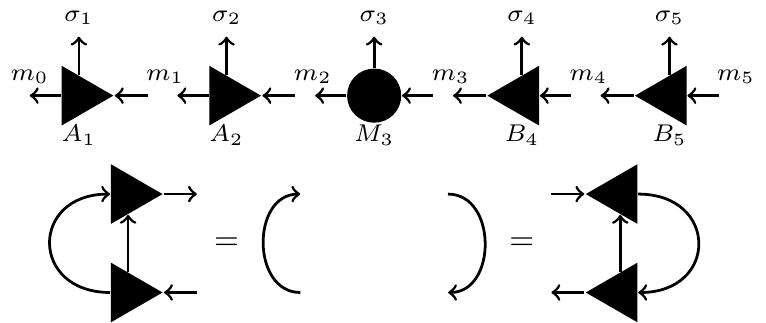}{
    \begin{tikzpicture}[baseline=-2]
      \node (a1) [regular polygon, regular polygon sides=3, shape border rotate=270, fill,minimum size=0.7cm] at (0,0){};
      \node (a1text) at (0,-0.4){\scriptsize{$A_1$}};
      \node (a2) [regular polygon, regular polygon sides=3, shape border rotate=270, fill,minimum size=0.7cm] at (1.5,0){};
      \node (a2text) at (1.5,-0.4){\scriptsize{$A_2$}};
      \node (m3) [circle, fill,minimum size=0.55cm] at (3,0){};
      \node (m3text) at (3,-0.4){\scriptsize{$M_3$}};
      \node (b4) [regular polygon, regular polygon sides=3, shape border rotate=90, fill,minimum size=0.7cm] at (4.5,0){};
      \node (b4text) at (4.5,-0.4){\scriptsize{$B_4$}};
      \node (b5) [regular polygon, regular polygon sides=3, shape border rotate=90, fill,minimum size=0.7cm] at (6,0){};
      \node (b5text) at (6,-0.4){\scriptsize{$B_5$}};
      
      \draw [thick,->] (a1) -- +(-0.5,0) node [left, above]{\scriptsize{$m_0$}};
      \draw [thick,<-] (a1) -- +(+0.7,0) node [right=5pt, above]{\scriptsize{$m_1$}};
      \draw [thick,->] (a1) -- +(0,0.6) node [above]{\scriptsize{$\sigma_1$}};

      \draw [thick,->] (a2) -- +(-0.5,0);
      \draw [thick,<-] (a2) -- +(+0.7,0) node [right=5pt, above]{\scriptsize{$m_2$}};
      \draw [thick,->] (a2) -- +(0,0.6) node [above]{\scriptsize{$\sigma_2$}};

      \draw [thick,->] (m3) -- +(-0.6,0);
      \draw [thick,<-] (m3) -- +(+0.6,0) node [right=5pt, above]{\scriptsize{$m_3$}};
      \draw [thick,->] (m3) -- +(0,0.6) node [above]{\scriptsize{$\sigma_3$}};

      \draw [thick,->] (b4) -- +(-0.7,0);
      \draw [thick,<-] (b4) -- +(+0.5,0) node [right=5pt, above]{\scriptsize{$m_4$}};
      \draw [thick,->] (b4) -- +(0,0.6) node [above]{\scriptsize{$\sigma_4$}};

      \draw [thick,->] (b5) -- +(-0.7,0);
      \draw [thick,<-] (b5) -- +(+0.5,0) node [right=5pt, above]{\scriptsize{$m_5$}};
      \draw [thick,->] (b5) -- +(0,0.6) node [above]{\scriptsize{$\sigma_5$}};

      \node (l1h) [regular polygon, regular polygon sides=3, shape border rotate=270, fill,minimum size=0.7cm] at (0.5,-1){};
      \node (l1) [regular polygon, regular polygon sides=3, shape border rotate=270, fill,minimum size=0.7cm] at (0.5,-2){};
      \draw [thick,->] (l1) .. controls +(left:+1cm) and +(left:1cm) .. (l1h);
      \draw [thick,->] (l1) -- (l1h);
      \draw [thick,<-] (l1) -- +(0.7,0);
      \draw [thick,->] (l1h) -- +(0.7,0);
      \node (eq1) at (1.5,-1.5){$=$};
      \draw [thick,->] (2.25,-2) .. controls +(left:+0.5cm) and +(left:0.5cm) .. (2.25,-1);

      \node (r1h) [regular polygon, regular polygon sides=3, shape border rotate=90, fill,minimum size=0.7cm] at (5.5,-1){};
      \node (r1) [regular polygon, regular polygon sides=3, shape border rotate=90, fill,minimum size=0.7cm] at (5.5,-2){};
      \draw [thick,<-] (r1) .. controls +(right:+1cm) and +(right:1cm) .. (r1h);
      \draw [thick,->] (r1) -- (r1h);
      \draw [thick,->] (r1) -- +(-0.7,0);
      \draw [thick,<-] (r1h) -- +(-0.7,0);
      \node (eq2) at (4.5,-1.5){$=$};
      \draw [thick,<-] (3.75,-2) .. controls +(right:+0.5cm) and +(right:0.5cm) .. (3.75,-1);

    \end{tikzpicture}
  }
  \caption{\label{fig:mps}Top: Graphical representation of a MPS in
    mixed-canonical form with the orthogonality center on site 3 and
    tensors $A_1, A_2, M_3, B_4, B_5$. Explicit tensor and tensor leg
    labels are given here. Bottom: Conditions for left and
    right-normalized tensors to result in identity matrices upon
    contraction. Labels are left off to avoid clutter.}
\end{figure}

\section{\label{sec:notation}MPS and MPO notation}

Matrix-product states (MPS) describe quantum mechanical states on a
separable Hilbert space $\mathcal{H} = \otimes_{i=1}^L \mathcal{H}_i$,
each with a local basis
$\left\{|\sigma_i\rangle\right\}_{\sigma_i=1}^{d_i}$. To represent a
state $|\psi\rangle$, $L$ rank-3 tensors
$M_{i; m_{i}}^{\sigma_i m_{i-1}}$ are selected such that
\begin{equation}
  |\psi\rangle = \sum_{\sigma_1} \ldots \sum_{\sigma_L} M_{1; m_1}^{\sigma_1 m_0} \cdots M_{L; m_L}^{\sigma_1  m_{L-1}} |\sigma_L \ldots \sigma_L \rangle \quad.
\end{equation}
Here, the $\cdot$ represents a contraction of the tensors over all
common indices, i.e.
\begin{equation}
  M_{1; m_1}^{\sigma_1 m_0} \cdot M_{2; m_2}^{\sigma_2 m_1} =
  \sum_{m_1} M_{1; m_1}^{\sigma_1 m_0} M_{2; m_2}^{\sigma_2 m_1} \quad.
\end{equation}
The $m_i$ are called \emph{right MPS bond indices}, $m_{i-1}$ are the
\emph{left MPS bond indices} and $\sigma_i$ are the \emph{local
  physical indices}. $m_0$ and $m_L$ are one-dimensional dummy indices
inserted for consistency. It is useful to differentiate between
incoming (lower, bra) and outgoing (upper, ket) indices in the context of
implementing symmetries in the network. An incoming index may only be
contracted with an outgoing index and vice-versa. Indices may be left
off if they are clear from context, e.g.:
\begin{equation}
  M_1 \cdot M_2 = \sum_{m_1} M_{1; m_1}^{\sigma_1 m_0} M_{2; m_2}^{\sigma_2 m_1} \quad.
\end{equation}
Furthermore, we will write $\boldsymbol{\sigma}$ to refer to all
$\sigma_1, \ldots, \sigma_L$.

Matrix-product state tensors may optionally be \emph{left-} or
\emph{right-normalized}. `A' instead of `M' will be used for
left-normalized tensors which fulfill
\begin{equation}
  \sum_{\sigma_i, m_{i-1}} A_{i; m_i}^{\sigma_i m_{i-1}} A_{i; \sigma_i m_{i-1}}^{\dagger; \tilde{m}_i} = \mathbf{1}_{m_i}^{\tilde{m}_i} \label{eq:leftnorm-matrix}
\end{equation}
and `B' will be used for right-normalized tensors fulfilling
\begin{equation}
  \sum_{\sigma_i, m_i} B_{i; m_i}^{\sigma_i m_{i-1}} B_{i; \sigma_i \tilde{m}_{i-1}}^{\dagger; m_i} = \mathbf{1}^{m_{i-1}}_{\tilde{m}_{i-1}} \quad,
\end{equation}
where ${}^\dagger$ denotes complex conjugation of all entries and
reversal of index directions such that
$A_{i; \sigma_i m_{i-1}}^{\dagger; \tilde{m}_i} \equiv
\big(A_i^\dagger \big)_{\sigma_i m_{i-1}}^{\tilde{m}_i} = \left[A_{i;
      m_i}^{\sigma_i m_{i-1}}\right]^\star$.
  A matrix-product state is in \emph{left-canonical}
  (\emph{right-canonical}) form if all tensors are left-normalized
  (right-normalized). A matrix-product state in which all tensors to
  the left of a specific site $k$ are left-normalized and all tensors
  to the right of that site $k$ are right-normalized is in
  \emph{mixed-canonical form}\cite{schollwoeck11} and site $k$ is its
  \emph{orthogonality center}\cite{stoudenmire10:_minim}
  (cf. Fig.~\ref{fig:mps}).

In a similar fashion, operators may be written as matrix-product
operators, consisting of $L$ rank-4 tensors
$W_{i; \sigma_i w_i}^{\tau_i w_{i-1}}$, such that:
\begin{equation}
  \hat H = \sum_{\boldsymbol{\sigma} \boldsymbol{\tau}} W_1 \cdot W_2 \cdots W_L |\boldsymbol{\tau} \rangle \langle \boldsymbol{\sigma} | \quad.
\end{equation}
Just like MPS tensors, MPO tensors have a \emph{left MPO bond index}
($w_{i-1}$), a \emph{right MPO bond index} ($w_i$) as well as an
\emph{upper} and \emph{lower physical index} ($\tau_i$ and $\sigma_i$
resp.). Multiple methods to construct MPOs from scratch exist.\cite{kin-lic16:_matrix_produc_operat_matrix_produc, hubig17:_gener, paeckel17:_autom}

\begin{figure}[b!]
  \centering
  \tikzpic{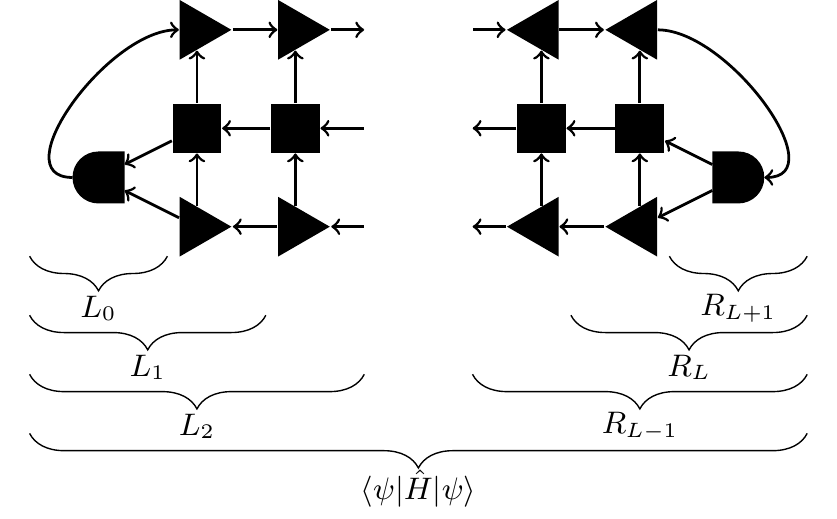}{
    \begin{tikzpicture}[baseline=-2]
      \node (f) [thick,draw, rounded rectangle, rounded rectangle right arc=none, fill, minimum size=0.5cm] at (-1,0.5){};
      \node (a1) [regular polygon, regular polygon sides=3, shape border rotate=270, fill,minimum size=0.7cm] at (0,0){};
      \node (a2) [regular polygon, regular polygon sides=3, shape border rotate=270, fill,minimum size=0.7cm] at (1,0){};
      \node (a1h) [regular polygon, regular polygon sides=3, shape border rotate=270, fill,minimum size=0.7cm] at (0,2){};
      \node (a2h) [regular polygon, regular polygon sides=3, shape border rotate=270, fill,minimum size=0.7cm] at (1,2){};
      \node (w1) [regular polygon, regular polygon sides=4, fill,minimum size=0.7cm] at (0,1){};
      \node (w2) [regular polygon, regular polygon sides=4, fill,minimum size=0.7cm] at (1,1){};

      \draw [thick,->] (a1) -- (f);
      \draw [thick,->] (w1) -- (f);
      \draw [thick,<-] (w1) -- (a1);
      \draw [thick,<-] (a1h) -- (w1);
      \draw [thick,->] (f) to [out=180,in=180] (a1h);
      \draw [thick,->] (a1h) -- (a2h);
      \draw [thick,->] (w2) -- (w1);
      \draw [thick,->] (a2) -- (a1);
      \draw [thick,<-] (w2) -- (a2);
      \draw [thick,<-] (a2h) -- (w2);
      \draw [thick,<-] (a2) -- +(+0.7,0);
      \draw [thick,<-] (w2) -- +(+0.7,0);
      \draw [thick,->] (a2h) -- +(+0.7,0);

      \draw [decorate, decoration={brace, amplitude=10pt, mirror},yshift=2pt] ( $ (f) + (-0.7, -0.8) $ ) -- ( $ (f) + (0.7, -0.8) $ ) node [black,midway,yshift=-15pt] {$L_0$};
      \draw [decorate, decoration={brace, amplitude=10pt, mirror},yshift=2pt] ( $ (f) + (-0.7, -1.4) $ ) -- ( $ (a1) + (0.7, -0.9) $ ) node [black,midway,yshift=-15pt] {$L_1$};
      \draw [decorate, decoration={brace, amplitude=10pt, mirror},yshift=2pt] ( $ (f) + (-0.7, -2) $ ) -- ( $ (a2) + (0.7, -1.5) $ ) node [black,midway,yshift=-15pt] {$L_2$};

      \node (s) [thick,draw, rounded rectangle, rounded rectangle left arc=none, fill, minimum size=0.5cm] at (5.5,0.5){};
      \node (b5) [regular polygon, regular polygon sides=3, shape border rotate=90, fill,minimum size=0.7cm] at (4.5,0){};
      \node (b4) [regular polygon, regular polygon sides=3, shape border rotate=90, fill,minimum size=0.7cm] at (3.5,0){};
      \node (b5h) [regular polygon, regular polygon sides=3, shape border rotate=90, fill,minimum size=0.7cm] at (4.5,2){};
      \node (b4h) [regular polygon, regular polygon sides=3, shape border rotate=90, fill,minimum size=0.7cm] at (3.5,2){};
      \node (w5) [regular polygon, regular polygon sides=4, fill,minimum size=0.7cm] at (4.5,1){};
      \node (w4) [regular polygon, regular polygon sides=4, fill,minimum size=0.7cm] at (3.5,1){};

      \draw [thick,->] (s) -- (b5);
      \draw [thick,->] (s) -- (w5);
      \draw [thick,<-] (w5) -- (b5);
      \draw [thick,<-] (b5h) -- (w5);
      \draw [thick,->] (b5h) to [out=0,in=0] (s);
      \draw [thick,->] (b4h) -- (b5h);
      \draw [thick,->] (w5) -- (w4);
      \draw [thick,->] (b5) -- (b4);
      \draw [thick,<-] (w4) -- (b4);
      \draw [thick,<-] (b4h) -- (w4);
      \draw [thick,->] (b4) -- +(-0.7,0);
      \draw [thick,->] (w4) -- +(-0.7,0);
      \draw [thick,<-] (b4h) -- +(-0.7,0);

      \draw [decorate, decoration={brace, amplitude=10pt, mirror},yshift=2pt] ( $ (s) + (-0.7, -0.8) $ ) --  ( $ (s) + (0.7, -0.8) $ ) node [black,midway,yshift=-15pt] {$R_{L+1}$};
      \draw [decorate, decoration={brace, amplitude=10pt, mirror},yshift=2pt] ( $ (b5) + (-0.7, -0.9) $ ) -- ( $ (s) + (0.7, -1.4) $ ) node [black,midway,yshift=-15pt] {$R_L$};
      \draw [decorate, decoration={brace, amplitude=10pt, mirror},yshift=2pt] ( $ (b4) + (-0.7, -1.5) $ ) -- ( $ (s) + (0.7, -2) $ ) node [black,midway,yshift=-15pt] {$R_{L-1}$};

      \draw [decorate, decoration={brace, amplitude=10pt, mirror},yshift=2pt] ( $ (f) + (-0.7, -2.6) $ ) --  ( $ (s) + (0.7, -2.6) $ ) node [black,midway,yshift=-16pt] {$\langle \psi | \hat H | \psi \rangle$};
    \end{tikzpicture}
  }
  \caption{\label{fig:contr}Consecutive left- and right-contractions
    of the MPO (squares) sandwiched between normalized MPS tensors
    (triangles). $L_0$ and $R_{L+1}$ are the dummy tensors inserted
    for consistency. Connecting any pair $L_i \cdot R_{i+1}$ results
    in the expectation value of the operator.}
\end{figure}
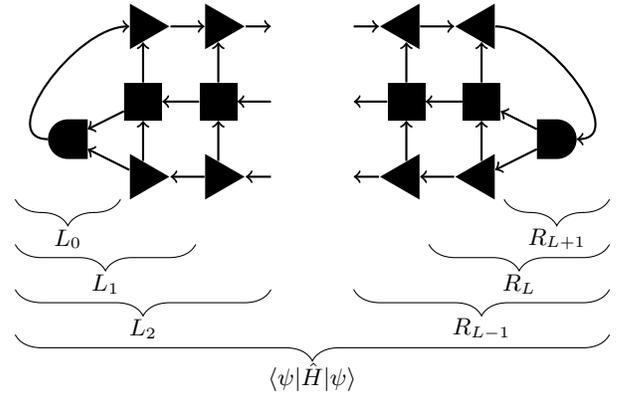

The following definitions will be useful later
(cf. Fig.~\ref{fig:contr}):
\begin{align}
  L_0 & = \mathbf{1}^{\tilde{m}_0}_{w_0 m_0} \\
  L_i & = L_{i-1} \cdot W_i \cdot A_i \cdot A_i^\dagger \\
  R_{L+1} & = \mathbf{1}^{w_L m_L}_{\tilde{m}_L} \\
  R_{i} & = R_{i+1} \cdot W_i \cdot B_i \cdot B_i^\dagger
\end{align}
The $L_i$ and $R_i$ are the usual left and right contractions of the
Hamiltonian sandwiched between the state as encountered during
standard DMRG, standard TDVP\cite{haegeman16:_unify} or expectation
value calculations. Hence, we can expect to be able to calculate them
efficiently. As a visualization, note that
\begin{align}
  L_i \cdot R_{i+1} = \langle \psi | \hat H | \psi \rangle \quad \forall i \in [0, L] \quad.
\end{align}
Throughout this paper, we will use $m$, $w$ and $d$ to denote the
effective MPS bond dimension, the effective MPO bond dimension and the
effective size of the local basis in particular when estimating the
computational cost of an operation.

Given an MPS tensor $A_i$, we can view it as an orthonormal basis
transformation and truncation from an effective left basis ($m_{i-1}$)
and local basis ($\sigma_i$) into a new effective right basis
($m_i$). However, the new right-hand side basis will typically not
span the complete space reachable from $m_{i-1} \otimes \sigma_i$, but
only a small subset thereof with size $m_i$. We hence define tensors
$F_i$ (and conversely $G_i$ when working with $B_i$) which reach the
additional $(d-1)m$ states with the properties
\begin{align}
  & \sum_{\sigma_i, m_{i-1}} A_{i; m_i}^{\sigma_i m_{i-1}} F_{i; \sigma_i m_{i-1}}^{\dagger; \tilde{m}^\prime_i} = 0 \label{eq:f-prop1}\\
  & \sum_{\sigma_i, m_{i-1}} F_{i; m^\prime_i}^{\sigma_i m_{i-1}} F_{i; \sigma_i m_{i-1}}^{\dagger; \tilde{m}^\prime_i} = \mathbf{1}_{m^\prime_i}^{\tilde{m}^\prime_i}
\end{align}
and equivalently for $G_i$:
\begin{align}
  & \sum_{\sigma_i, m_i} B_{i; m_i}^{\sigma_i m_{i-1}} G_{i; \sigma_i \tilde{m}^\prime_{i-1}}^{\dagger; m_i} = 0 \\
  & \sum_{\sigma_i, m_i} G_{i; m_i}^{\sigma_i m^\prime_{i-1}} G_{i; \sigma_i \tilde{m}^\prime_{i-1}}^{\dagger; m_i} = \mathbf{1}_{\tilde{m}^\prime_{i-1}}^{m^\prime_{i-1}} \label{eq:g-prop2}
\end{align}
When interpreting $A_{i; m_i}^{\sigma_i m_{i-1}}$ as a rectangular
matrix whose row index is obtained from joining the left virtual and
physical index (i.e. the two upper indices), it is an isometric matrix
in the sense of Eq.~\eqref{eq:leftnorm-matrix}. Similarly interpreting
$F_{i; m^\prime_i}^{\sigma_i m_{i-1}}$, it corresponds to the
additional columns required to extend the isometric matrix $A_i$ into
a square unitary matrix. Put differently, if $A_i$ is obtained from a
`reduced' or `thin' QR decomposition, one can similarly obtain $F_i$
by instead requesting a `full' decomposition\footnote{Care must be
  taken such that the full QR decomposition also considers valid zero
  blocks of $A_i$ if such zero blocks are not stored by the tensor
  library. It may be helpful to first construct a full isometry
  mapping the left MPS basis and the local physical basis into a
  maximal right MPS basis, then resize the right-hand side blocks of
  this isometry to be compatible with $A_i$, multiply the entire
  isometry by a tiny number and add it to $A_i$ prior to the QR
  decomposition. This way, the QR also `sees' all possible zero
  blocks.} with a square and unitary Q matrix of size $md \times md$
(assuming that the left bond dimension is $m$). The first $m$ columns
correspond to $A_{i; m_i}^{\sigma_i m_{i-1}}$, whereas the last
$m (d-1)$ columns define $F_{i; m^\prime_i}^{\sigma_i m_{i-1}}$, with
thus $m^\prime_i = 1,\ldots,m(d-1)$. An analogous construction defines
$G_{i; m_i}^{\sigma_i m^\prime_{i-1}}$.

\section{\label{sec:alternatives}Current alternatives}

\subsection{\label{sec:alternatives:2dmrg}2DMRG truncation error}

The 2DMRG truncation error is readily available from a 2DMRG
calculation and has repeatedly been shown\cite{legeza96:_accur,
  white05:_densit, leblanc15:_solut_two_dimen_hubbar_model,
  ehlers17:_hybrid_hubbar} to allow a reliably extrapolation of
observables obtained during the calculation towards the
infinite-precision ground state. In the examples later in the paper,
we have taken the largest 2DMRG truncation error and the lowest
eigensolver energy encountered during the last half-sweep at a given
bond dimension as error measure and expectation value respectively.

However, the 2DMRG method both scales relatively badly in the local
physical dimension as $O(m^3 d^2 w + m^2 d^3 w^2 + m^3 d^3)$ and
sometimes -- in particular, if no noise terms are used -- is slow to
pick up long-range correlations.\cite{white05:_densit} Generally, one
can expect a speed-up of approximately four in fermionic or spin
systems when switching to a single-site implementation to obtain the
same accuracy in energy.\cite{hubig15:_stric_dmrg} Furthermore, the
idea of updating two sites at the same time runs somewhat counter to
the original aim of matrix-product states, namely reducing the
exponential complexity of the Hilbert space as much as
possible. Finally, the 2DMRG truncation error is obtained during the
2DMRG calculation and hence applies to the DMRG process itself, not
necessarily to the resulting state. If the state is already
well-converged at the current bond dimension and hence changes little
in subsequent sweeps, the difference will be minimal and the
extrapolation can be applied correctly. However, it may be difficult
to pinpoint this convergence during a large-scale calculation.

\subsection{\label{sec:alternatives:variance}The full variance}

Evaluation of the full variance
$\langle \psi | \hat H^2 | \psi \rangle - \langle \psi | \hat H | \psi
\rangle^2$
directly yields information on the non-eigenstate content of the
(assumed normalized) state $|\psi\rangle$: With
$E = \langle \psi | \hat H | \psi \rangle$, the residuum
$|\phi\rangle$ is given as
\begin{align}
  |\phi\rangle & = \hat H |\psi\rangle - E |\psi\rangle \\
  \Rightarrow \langle \phi | \phi \rangle & = \langle \psi | \hat H^2 | \psi \rangle - E^2 \quad.
\end{align}
The variance of $\hat H$ with respect to our current state
$|\psi\rangle$ is hence the norm squared of the residuum
$|\phi\rangle = \hat H |\psi\rangle - E|\psi\rangle$. It is an
extremely useful tool to check the convergence of DMRG and -- contrary
to the 2DMRG truncation error -- can not just diagnose insufficient
bond dimensions but also other convergence problems. When
extrapolating the energy $E$ in the variance $v$, we typically expect
a linear behavior, i.e.~$E(v) = a \cdot v + E_0$. For other
observables, the exponent may be different from one depending on the
system at hand, the observable and how well either 1DMRG or 2DMRG can
optimize this observable. In Sec.~\ref{sec:examples:hubbard}, we
provide one example to show the different range of exponents
potentially encountered in such extrapolations.

Unfortunately, calculation of $\langle \hat H^2 \rangle$ is
computationally relatively expensive. A naive evaluation scales as
$O(m^3 d w^2 + m^2 d^2 w^3)$. First evaluating $\hat H^2$ and applying
a MPO compression scheme\cite{froewis10:_tensor, hubig17:_gener}
allows us to reduce this to the calculation of the expectation value
of a larger MPO with bond dimension $w^\prime$. In most cases,
$w^\prime \approx 2 w$. While this is unproblematic for simple
one-dimensional systems with nearest-neighbor interactions, more
complicated systems (e.g.~from embedded problems, cylindrical systems
or direct application of DMRG to quantum chemistry models) also result
in much larger MPO bond dimensions $w$ which make evaluation of the
variance unfeasible or at least much more costly than the initial DMRG
calculation which lead to the state $|\psi\rangle$.  In particular,
there is a region (roughly $m \approx 10000$ and $w \approx 50$) where
DMRG calculations and evaluation of simple observables are possible,
but evaluating $\langle \psi | \hat H^2 |\psi\rangle$ or
$\hat H |\psi\rangle$ is not.

\section{\label{sec:prop}Proposed new error measure: 2-site variance}

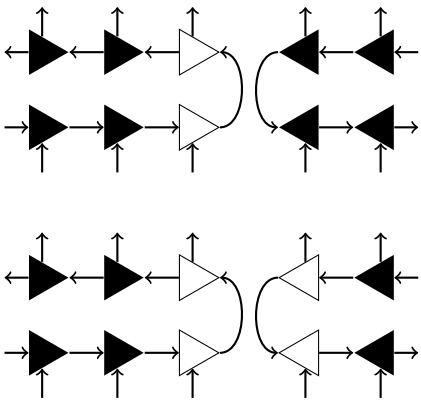
\begin{figure}[b]
  \centering
  \tikzpic{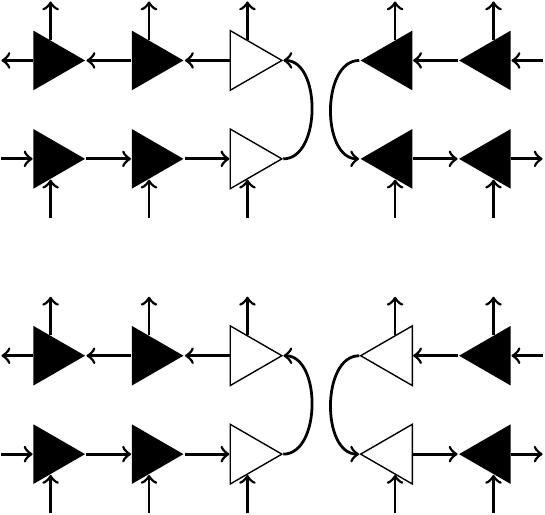}{
    \begin{tikzpicture}[baseline=-2]
      \node (a1) [regular polygon, regular polygon sides=3, shape border rotate=270, fill,minimum size=0.7cm] at (0,0){};
      \node (a2) [regular polygon, regular polygon sides=3, shape border rotate=270, fill,minimum size=0.7cm] at (1,0){};
      \node (f3) [regular polygon, regular polygon sides=3, shape border rotate=270, draw, minimum size=0.7cm] at (2,0){};
      \node (b4) [regular polygon, regular polygon sides=3, shape border rotate=90, fill,minimum size=0.7cm] at (3.5,0){};
      \node (b5) [regular polygon, regular polygon sides=3, shape border rotate=90, fill,minimum size=0.7cm] at (4.5,0){};

      \node (a1l) [regular polygon, regular polygon sides=3, shape border rotate=270, fill,minimum size=0.7cm] at (0,-1){};
      \node (a2l) [regular polygon, regular polygon sides=3, shape border rotate=270, fill,minimum size=0.7cm] at (1,-1){};
      \node (f3l) [regular polygon, regular polygon sides=3, shape border rotate=270, draw, minimum size=0.7cm] at (2,-1){};
      \node (b4l) [regular polygon, regular polygon sides=3, shape border rotate=90, fill,minimum size=0.7cm] at (3.5,-1){};
      \node (b5l) [regular polygon, regular polygon sides=3, shape border rotate=90, fill,minimum size=0.7cm] at (4.5,-1){};

      \draw [thick,<-] (a1) -- (a2);
      \draw [thick,->] (a1) -- +(-0.5,0);
      \draw [thick,->] (a1) -- +(0,0.6);
      \draw [thick,<-] (a2) -- (f3);
      \draw [thick,->] (a2) -- +(0,0.6);
      \draw [thick,->] (f3) -- +(0,0.6);
      \draw [thick,<-] (b4) -- (b5);
      \draw [thick,->] (b4) -- +(0,0.6);
      \draw [thick,->] (b5) -- +(0,0.6);
      \draw [thick,<-] (b5) -- +(0.5,0);
      \draw [thick,->] (f3l) .. controls +(right:+0.75cm) and +(right:0.75cm) .. (f3);
      \draw [thick,<-] (b4l) .. controls +(left:+0.75cm) and +(left:0.75cm) .. (b4);

      \draw [thick,->] (a1l) -- (a2l);
      \draw [thick,<-] (a1l) -- +(-0.5,0);
      \draw [thick,<-] (a1l) -- +(0,-0.6);
      \draw [thick,->] (a2l) -- (f3l);
      \draw [thick,<-] (a2l) -- +(0,-0.6);
      \draw [thick,<-] (f3l) -- +(0,-0.6);
      \draw [thick,->] (b4l) -- (b5l);
      \draw [thick,<-] (b4l) -- +(0,-0.6);
      \draw [thick,<-] (b5l) -- +(0,-0.6);
      \draw [thick,->] (b5l) -- +(0.5,0);

      \node (a1) [regular polygon, regular polygon sides=3, shape border rotate=270, fill,minimum size=0.7cm] at (0,-3){};
      \node (a2) [regular polygon, regular polygon sides=3, shape border rotate=270, fill,minimum size=0.7cm] at (1,-3){};
      \node (f3) [regular polygon, regular polygon sides=3, shape border rotate=270, draw, minimum size=0.7cm] at (2,-3){};
      \node (f4) [regular polygon, regular polygon sides=3, shape border rotate=90, draw, minimum size=0.7cm] at (3.5,-3){};
      \node (b5) [regular polygon, regular polygon sides=3, shape border rotate=90, fill,minimum size=0.7cm] at (4.5,-3){};

      \node (a1l) [regular polygon, regular polygon sides=3, shape border rotate=270, fill,minimum size=0.7cm] at (0,-4){};
      \node (a2l) [regular polygon, regular polygon sides=3, shape border rotate=270, fill,minimum size=0.7cm] at (1,-4){};
      \node (f3l) [regular polygon, regular polygon sides=3, shape border rotate=270, draw, minimum size=0.7cm] at (2,-4){};
      \node (f4l) [regular polygon, regular polygon sides=3, shape border rotate=90, draw, minimum size=0.7cm] at (3.5,-4){};
      \node (b5l) [regular polygon, regular polygon sides=3, shape border rotate=90, fill,minimum size=0.7cm] at (4.5,-4){};

      \draw [thick,<-] (a1) -- (a2);
      \draw [thick,->] (a1) -- +(-0.5,0);
      \draw [thick,->] (a1) -- +(0,0.6);
      \draw [thick,<-] (a2) -- (f3);
      \draw [thick,->] (a2) -- +(0,0.6);
      \draw [thick,->] (f3) -- +(0,0.6);
      \draw [thick,<-] (f4) -- (b5);
      \draw [thick,->] (f4) -- +(0,0.6);
      \draw [thick,->] (b5) -- +(0,0.6);
      \draw [thick,<-] (b5) -- +(0.5,0);
      \draw [thick,->] (f3l) .. controls +(right:+0.75cm) and +(right:0.75cm) .. (f3);
      \draw [thick,<-] (f4l) .. controls +(left:+0.75cm) and +(left:0.75cm) .. (f4);

      \draw [thick,->] (a1l) -- (a2l);
      \draw [thick,<-] (a1l) -- +(-0.5,0);
      \draw [thick,<-] (a1l) -- +(0,-0.6);
      \draw [thick,->] (a2l) -- (f3l);
      \draw [thick,<-] (a2l) -- +(0,-0.6);
      \draw [thick,<-] (f3l) -- +(0,-0.6);
      \draw [thick,->] (f4l) -- (b5l);
      \draw [thick,<-] (f4l) -- +(0,-0.6);
      \draw [thick,<-] (b5l) -- +(0,-0.6);
      \draw [thick,->] (b5l) -- +(0.5,0);
    \end{tikzpicture}
  }
  \caption{\label{fig:projectors}Top: Individual term of the projector
    $\hat P_1$, to be summed over all sites. Bottom: Individual term
    of the projector $\hat P_2$, to be summed over all pairs of
    neighboring sites. Tensors $F_i$ and $G_i$ are drawn as
    non-filled triangles.}
\end{figure}

The complete Hilbert space $\mathcal{H}$ may not only be decomposed
into a product of local Hilbert spaces, but, given a MPS
$|\psi\rangle$, also into a direct sum of orthogonal spaces
\begin{equation}
  \mathcal{H} = \bigoplus_{l=0}^L \mathcal{W}_l
\end{equation}
where $\mathcal{W}_0$ is the one-dimensional space of states parallel
to $|\psi\rangle$ and $\mathcal{W}_l$ are the spaces of variations of
$l$ continuous sites orthogonal to all $\mathcal{W}_{k < l}$.
$\mathcal{W}_L$ could potentially span the entirety of $\mathcal{H}$
due to the completeness of matrix-product states, with only the
subspaces already contained in $\mathcal{W}_{k<L}$ removed from
it. Depending on the state $|\psi\rangle$, the partition of
$\mathcal{H}$ into $\mathcal{W}_l$ changes.

Specifically, $\mathcal{W}_1$ is spanned by the states
\begin{align}
  & |\phi^{(1)}_i(V)\rangle = \sum_{\boldsymbol{\sigma}} \cdots A_{i-1} \cdot F_i \cdot V_i \cdot B_{i+1} \cdots | \boldsymbol{\sigma}\rangle \nonumber \\
  & \quad \forall i \in [1, L] \quad \forall V_i \in \mathbb{C}^{m_i^\prime}_{m_i}
\end{align}
and similarly $\mathcal{W}_2$ by the states
\begin{align}
  & |\phi^{(2)}_{i,i+1}(W)\rangle = \sum_{\boldsymbol{\sigma}} \cdots A_{i-1} \cdot F_i \cdot W_i \cdot G_{i+1} \cdot B_{i+2} \cdots | \boldsymbol{\sigma}\rangle \nonumber \\
  & \quad \forall i \in [1, L-1] \quad \forall W_i \in \mathbb{C}^{m_i^\prime}_{m^\prime_i} \quad.
\end{align}
The tensors $F_i$ and $G_i$ have the properties as defined in
Eqs.~\eqref{eq:f-prop1} through \eqref{eq:g-prop2}.  The projector
$\hat P_1$ into the space $\mathcal{W}_1$ is given by
$\sum_i |\phi^{(1)}_i\rangle\langle\phi^{(1)}_i|$ with the matrix
$V_i$ left off and the left-hand side legs of $B_{i+1}$ and
$B^{\dagger}_{i+1}$ as well as the right-hand side legs of $F_i$ and
$F^\dagger_i$ connected. Similarly, the projector $\hat P_2$ into the
space $\mathcal{W}_2$ is given by
$\sum_i |\phi^{(2)}_{i,i+1}\rangle\langle\phi^{(2)}_{i,i+1}|$
connected in the same way. Individual terms of these two projectors
are illustrated in Fig.~\ref{fig:projectors}. The projector $\hat P_0$
for $\mathcal{W}_0$ is simply $|\psi\rangle\langle\psi|$.

If we now consider the full variance
$\langle \psi | (\hat H - E) (\hat H - E)|\psi\rangle$ of a normalized
state $|\psi\rangle$, we may insert an identity
$\mathbf{1} = \sum_{l=0}^L \hat P_l$:
\begin{align}
           & \langle \psi | (\hat H - E) (\hat H - E)|\psi\rangle \\
  =        & \langle \psi | (\hat H - E) \left[ \sum_{l=0}^L \hat P_l \right] (\hat H - E)|\psi\rangle \\
  \approx & \langle \psi | (\hat H - E) \left(\hat P_0 + \hat P_1 + \hat P_2\right) (\hat H - E)|\psi\rangle \label{eq:approx} \\
  =       & \langle \psi | \hat H \hat P_1 \hat H |\psi\rangle + \langle \psi | \hat H \hat P_2 \hat H |\psi\rangle \quad,  \label{eq:p1p2}
\end{align}
where all other terms of the form
$\langle \psi | \hat H \hat P_{1,2} E | \psi\rangle$ are identically
zero due to the orthogonality of $\hat P_{1,2}|\phi\rangle$ and
$|\psi\rangle$ for all $|\phi\rangle$ and
$\langle \psi|(\hat H - E) \hat P_0 (\hat H - E) |\psi\rangle \equiv 0$ as
well.

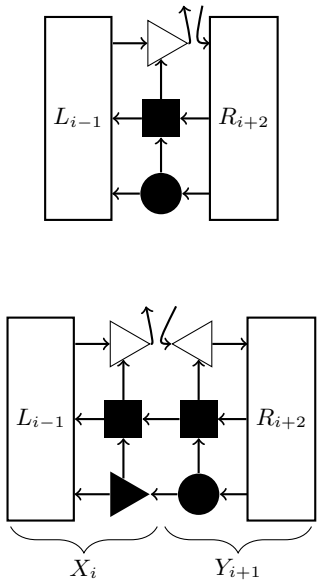
\begin{figure}[b]
  \centering
  \tikzpic{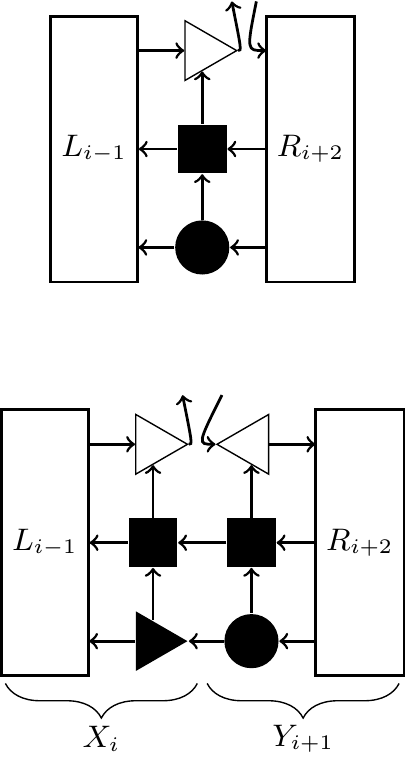}{
    \begin{tikzpicture}[baseline=-2]
      \node (l) [draw,thick,rectangle, minimum height=2.7cm, minimum width=0.7cm] at (0.4,1){$L_{i-1}$};
      \node (a1) [fill,circle,minimum size=0.55cm] at (1.5,0){};
      \node (f1) [regular polygon, regular polygon sides=3, shape border rotate=270, draw, minimum size=0.7cm] at (1.5,2){};
      \node (w1) [regular polygon, regular polygon sides=4, fill,minimum size=0.7cm] at (1.5,1){};

      \draw [thick,<-] (w1) -- (a1);
      \draw [thick,->] (w1) -- (l);
      \draw [thick,->] (a1) -- +(-0.65,0);
      \draw [thick,->] (0.86,2) -- (f1);
      \draw [thick,->] (w1) -- (f1);
      \draw [thick,->] (f1) .. controls +(right:+0.4cm) .. +(0.3,0.5);

      \node (r) [rectangle, draw, thick, minimum height=2.7cm, minimum width=0.7cm] at (2.6,1){$R_{i+2}$};
      \draw [thick,<-] (2.15,2) .. controls +(left:+0.2cm) .. +(-0.1,0.5);

      \draw [thick,<-] (a1) -- +(0.65,0);
      \draw [thick,<-] (w1) -- (r);
    
      \node (l) [draw,thick,rectangle, minimum height=2.7cm, minimum width=0.7cm] at (-0.1,-3){$L_{i-1}$};
      \node (a1) [regular polygon, regular polygon sides=3, shape border rotate=270, fill, minimum size=0.7cm] at (1,-4){};
      \node (f1) [regular polygon, regular polygon sides=3, shape border rotate=270, draw, minimum size=0.7cm] at (1,-2){};
      \node (w1) [regular polygon, regular polygon sides=4, fill,minimum size=0.7cm] at (1,-3){};

      \draw [thick,<-] (w1) -- (a1);
      \draw [thick,->] (w1) -- (l);
      \draw [thick,->] (a1) -- +(-0.65,0);
      \draw [thick,->] (0.36,-2) -- (f1);
      \draw [thick,->] (w1) -- (f1);
      \draw [thick,->] (f1) .. controls +(right:+0.4cm) .. +(0.3,0.5);

      \node (r) [rectangle, draw, thick, minimum height=2.7cm, minimum width=0.7cm] at (3.1,-3){$R_{i+2}$};
      \node (m2) [fill,circle,minimum size=0.55cm] at (2,-4){};
      \node (w2) [regular polygon, regular polygon sides=4, fill,minimum size=0.7cm] at (2,-3){};
      \node (f2) [regular polygon, regular polygon sides=3, shape border rotate=90, draw, minimum size=0.7cm] at (2,-2){};
      \draw [thick,<-] (f2) .. controls +(left:+0.55cm) .. +(-0.3,0.5);

      \draw [thick,<-] (w2) -- (m2);
      \draw [thick,<-] (w2) -- (r);
      \draw [thick,<-] (m2) -- +(0.65,0);
      \draw [thick,->] (f2) -- +(0.65,0);
      \draw [thick,->] (w2) -- (f2);

      \draw [thick,->] (w2) -- (w1);
      \draw [thick,->] (m2) -- (a1);

      \draw [decorate, decoration={brace, amplitude=10pt, mirror},yshift=2pt] (-0.5,-4.5) -- (1.45,-4.5) node [black,midway,yshift=-16pt] {$X_i$};
      \draw [decorate, decoration={brace, amplitude=10pt, mirror},yshift=2pt] (1.55,-4.5) -- (3.5,-4.5) node [black,midway,yshift=-16pt] {$Y_{i+1}$};
    \end{tikzpicture}
    }
    \caption{\label{fig:terms}Top: One of the $L$ one-site
      contributions to the variance. Bottom: One of the $L-1$ two-site
      contributions to the variance. Temporary tensors $X_i$ and
      $Y_{i+1}$ (excluding $F_i$ and $G_{i+1}$), to be calculated
      during the evaluation of $L_i$ an $R_{i+1}$, are marked. Tensors
      $F_i$ and $G_i$ are drawn as non-filled triangles.}
\end{figure}

There are a total of $2L - 1$ terms in Eq.~\eqref{eq:p1p2}, all of
which can be written as squared Frobenius norms of rank-2 tensors
(cf.~Fig.~\ref{fig:terms}).

Note that, if the Hamiltonian is composed of nearest-neighbor
interactions only, the approximation in Eq.~\eqref{eq:approx} becomes
an equality. In this case, the Hamiltonian is a sum of
nearest-neighbor terms $\hat h_{i,i+1}$. Applying such a term to the
state $|\psi\rangle$ only has to change its MPS tensors on sites $i$
and $i+1$ (to see this, consider $\hat h_{i,i+1}$ as a two-site MPO
gate). $\hat h_{i,i+1} |\psi\rangle$ is hence contained in
$\mathrm{span}\left(|\psi\rangle, |\phi^{(1)}_i(V)\rangle,
  |\phi^{(2)}_{i,i+1}(W)\rangle\right)$
for suitably-chosen $V, W$ and hence in
$\mathcal{W}_0 \oplus \mathcal{W}_1 \oplus \mathcal{W}_2$.

Equally, if we were to include also $\hat P_3$, we could
calculate the variance of a three-site operator exactly (albeit at
$d$-times higher computational effort).

\begin{figure}
  \centering
  \includegraphics[width=\columnwidth]{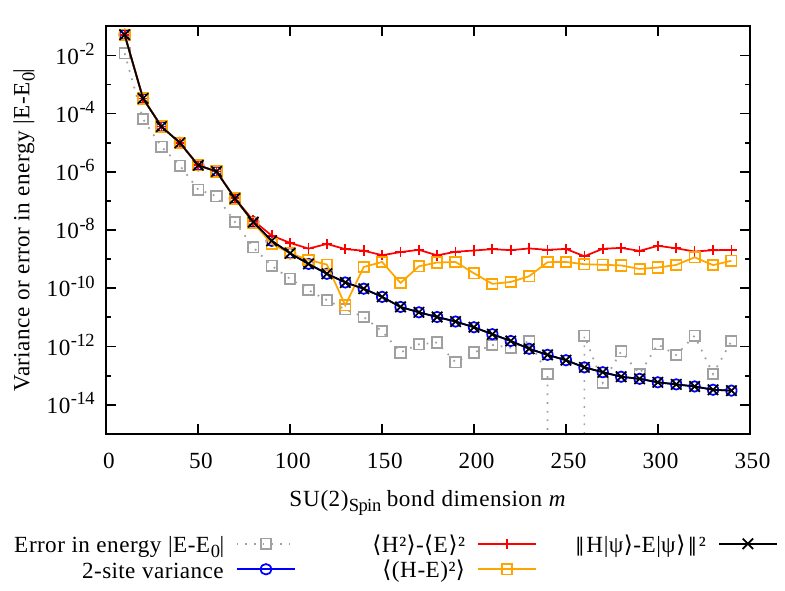}
  \caption{\label{fig:precise-variance}(Color online) Comparison of
    different ways to evaluate the variance
    $\langle (\hat H - E)^2 \rangle$ of a $L=200$ sites $S=1$
    Heisenberg model at different
    $\mathrm{SU}(2)_{\mathrm{Spin}}$-invariant bond dimensions. The
    error in energy relative to
    $E_0 \approx -279.088 490 029 140 \pm 2 \times 10^{-12}$ is given
    in gray for comparison and shown to saturate numerical precision around
    $m \approx 180$. Of the three alternatives to the 2-site variance,
    only
    $|| \hat H |\psi\rangle - E|\psi\rangle
      ||^2$ is equally precise but requires a full MPO-MPS
    product and an MPS-MPS addition, making it much more costly.}
\end{figure}

In such cases, the 2-site variance proposed here is actually a
remarkably stable and numerically precise way to evaluate the
variance: the large terms of order $E^2$ are removed exactly, which
would otherwise incur a loss of approximately $\mathrm{log}_{10}(E^2)$
digits of precision when evaluating
$\langle H^2 \rangle - \langle E \rangle^2$ directly. We also avoid
the alternative subtraction in $\langle (\hat H - E)^2\rangle$ which
still incurs losing approximately $\mathrm{log}_{10}(|E|)$
digits. Instead, only positive, small terms of order
$\langle (H - E)^2 \rangle / L$ are added together. We are left with
the unavoidable loss of precision due to repeated matrix-matrix
products of one or two digits relative to the machine epsilon. This
effect is demonstrated in Fig.~\ref{fig:precise-variance}, where four
possible approaches to evaluate the variance in a $S=1$, $L=200$
Heisenberg chain with open boundary conditions are compared. The
2-site variance is one of the two most precise methods and also the
fastest method: for example, the last data points at $m = 340$ in
Fig.~\ref{fig:precise-variance} took \unit{26}{s} for the 2-site
variance, \unit{41}{s} for
$\langle \hat H^2 \rangle - \langle E \rangle^2$, \unit{41}{s} for
$\langle (\hat H - E)^2 \rangle$ and \unit{436}{s} for
$\left|\left| \hat H |\psi\rangle - E|\psi\rangle \right|\right|^2$ on
a two-core Intel i5-6200U CPU. This system is the best-case scenario
for calculation of the full variance due to the small original bond
dimension of $\hat H$ with just $w = 5$. On more complicated systems,
the relative advantage of the 2-site variance will be more pronounced.

Depending on the number of cores available, different procedures yield
the fastest wall-clock time and least memory/temporary disk space usage when evaluating the 2-site variance:

If only a single core is available, it is reasonable to first
left-normalize $|\psi\rangle$, evaluate $L_{i}$ and store
$X_{i}\cdot F_{i}^\dagger$ (cf.~Fig.~\ref{fig:terms},
$X_i = L_{i-1} \cdot W_i \cdot A_i$) for all sites. The $L_i$ do not
have to be stored. Then, sweeping right-to-left, one first
right-normalizes $M_{i+1}$ into $B_{i+1}$ and $T$. $B_{i+1}$ is used
to evaluate $Y^\prime_{i+1} = R_{i+2} \cdot W_{i+1} \cdot B_{i+1}$
which is stored temporarily. $G^\dagger_{i+1}$ and $T$ are contracted
into $Y^\prime_{i+1}$ and the result is then contracted with the left
half to yield the rank-2 tensor depicted in Fig.~\ref{fig:terms},
bottom panel. The squared Frobenius norm of this tensor is taken and
added to the accumulator. One then evaluates $R_{i+1}$ by re-using
$Y^\prime_{i+1}$, moves $T$ into the next site tensor to the left as
well as into the contraction $X_{i}$ to yield
$X^\prime_{i}$. $X^\prime_{i}$ is contracted with $R_{i+1}$ to give
the tensor in Fig.~\ref{fig:terms}, upper panel. Its squared Frobenius
norm is again added to the accumulator and one moves to the next site.

If, on the other hand, many cores are available, it is reasonable to
parallelize the most expensive part, namely the calculation of $F_i$
and $G_i$ as well as the products leading to the tensors in
Fig.~\ref{fig:terms}. This can be done by first evaluating $X_i$ and
$Y_i$ as well as $R_i$ on all sites by two independent processes acting
on two left- and right-normalized copies of $|\psi\rangle$. Once these
contractions are available and e.g.~stored temporarily on disk, one
may start $2L-1$ processes, each evaluating one of the $2L-1$ individual
terms.

The costs of this procedure are distributed as follows: Left- and
right-normalization as well as calculation of $X_i$, $Y_i$, $R_i$ and
the (temporary) $L_i$ all scale as $O(m^3 d w)$ and are roughly twice
as expensive as calculating a single expectation value
$\langle \psi | \hat H | \psi \rangle$, but can be parallelized
two-fold. The generation\footnote{To achieve the optimal scaling of
  $O(m^3 d(d-1))$, one needs to multiply an appropriately-formed
  $F^\prime$ matrix into the packed \textsc{LAPACK} representation of
  $Q$ to directly return the desired matrix form of $F$ of size
  $m \times m (d-1)$. Alternatively first constructing the full $Q$
  matrix and then removing the first few columns requires cost
  $O(m^3 d^2)$.} of $F_i$ and $G_i$ scales as $O(m^3 d(d-1))$ but can
be parallelized $2L-1$-fold. Contractions $X_i \cdot F^\dagger_i$ and
$Y_{i+1} \cdot G_{i+1}^\dagger$ cost $O(m^3 d (d-1) w)$ each, the
contraction $X_i F_i^\dagger \cdot Y_{i+1} G_{i+1}^\dagger$ costs
$O(m^3 (d-1)^2 w)$, but these can also be parallelized perfectly.

As such, the serial part of the calculation takes wall-clock time
comparable to a single expectation value calculation. The following,
$2L-1$-fold parallelized part scales worse in the local physical
dimension than the pure 1DMRG, but already better than 2DMRG. Its
primary components, the two full QRs to calculate $F_i$ and $G_i$, are
also much cheaper than the SVD of the two-site tensor in 2DMRG both
asymptotically (by a factor of $d$) and in practical calculations.

\section{\label{sec:examples}Examples}

The first two examples are intended to show that the variance itself
is a valid extrapolation tool and as useful as the 2DMRG truncation
error. This is done at the example of nearest-neighbor interaction
chains of Heisenberg spins and Hubbard electrons. Next, we show at the
example of long-range Coulomb-like interactions that even if the full
variance and two-site variance do not coincide due to long-range
interactions, both yield comparable results. Finally, we consider the
Hubbard model on a cylinder where it is impractical to calculate the
full variance but extrapolation in the two-site variance is as useful
as extrapolation in the 2DMRG truncation error.

\subsection{\label{sec:examples:intro}Introductory remarks}

In the following Figs.~\ref{fig:examples:heisenberg} through
\ref{fig:examples:hubbard-2d-zoom}, we show stages of the calculations
always as points with select bond dimensions indicated by nearby
numbers. The y-axis position of each point is given by the observable
expectation value (in Figs.~\ref{fig:examples:hubbard-2d} and
\ref{fig:examples:hubbard-2d-zoom}) or the error compared to the true
ground state expectation value (in
Figs.~\ref{fig:examples:heisenberg}-\ref{fig:examples_longrange}) at
that particular stage. The x-axis position is given by the error
measurement used, i.e.  either the 2DMRG truncation error (always in
green), the full variance (in red where available) or the 2-site
variance (always in blue) as observed at this stage.  Errors in energy
and error measures smaller than the plot range (typically $10^{-14}$)
were clipped to that value for illustrative purposes. The error
measures were likewise scaled by constant factors to fit into the same
plot (this does not affect the extrapolation).

For the energies plotted in Figs.~\ref{fig:examples:heisenberg},
\ref{fig:examples:hubbard} and
\ref{fig:examples_longrange}-\ref{fig:examples:hubbard-2d-zoom},
linear extrapolations towards zero error were attempted over several
intervals, each containing a certain number of data points obtained
from the calculation. In the plots, the least to most accurate
extrapolations are always shown as dotted, dashed, dash-dotted and
solid lines respectively. In Fig.~\ref{fig:examples:hubbard:n}, the
exponent was also selected as a fit parameter and only one
extrapolation performed per data set.

Ideally, extrapolations over intervals with smaller bond dimensions are
validated by calculations at higher bond dimensions: In Fig.~\ref{fig:examples:heisenberg}, we
would like the data points at bond dimension 160 to lie on the line
extrapolated from bond dimensions $[2,\ldots,m^\prime \ll 160]$. We can
hence also judge the quality of an extrapolation by observing its
change when including additional data points.

Furthermore, a correct extrapolation in
Figs.~\ref{fig:examples:heisenberg} through
\ref{fig:examples_longrange} would result in a y-intercept of
zero, indicating that the extrapolation produced the exact
(error-free) value. Deviations from this ideal case (i.e., non-zero
y-intercepts) result in saturated constant extrapolations at small
error values. If the extrapolated value is smaller than the true
ground-state value, the resulting zero crossing displays as a narrow
dip in the extrapolation curve.

In Fig.~\ref{fig:examples:hubbard-2d} and
\ref{fig:examples:hubbard-2d-zoom}, no exact reference values are
available, making the log-log plot impossible. We hence show in total
three different ranges of the calculation with linear scales on both
axes.

For the 2DMRG calculations, the energy is measured during the
calculation and prior to each local truncation, resulting in a larger
effective bond dimension for 2DMRG and a slightly lower energy. This
would also be done in actual calculations specifically to exploit this
locally larger dimension for more accurate measurements. We hence have
not eliminated this advantage by bringing the 2DMRG state into
canonical form before evaluating its energy. For the variance
measures, 1DMRG with subspace expansion
(DMRG3S\cite{hubig15:_stric_dmrg}) is used and the energy is evaluated
as the expectation value of the Hamiltonian with respect to the final
MPS.

\subsection{\label{sec:examples:heisenberg}Heisenberg spin chain}

As the first example, we wish to analyze the convergence
behavior of a $L=30$ Heisenberg spin chain with open boundary
conditions. Only the $\mathrm{U}(1)$ symmetry is implemented, but the
Hamiltonian itself is rotationally invariant:
\begin{equation}
  \hat H = \sum_{i=1}^{29} \sum_{a=x,y,z} \hat s^a_i \hat s^a_{i+1} \quad.
\end{equation}
30 sweeps each are run with bond dimensions
$m = 2, 4, 6, 8, \ldots, 160$. For the reference value of the
ground-state energy $E_0 = -13.1113557586032048\pm 5\times 10^{-14}$,
a calculation with $m = 500$ is run. This ground state can be
truncated with a truncation error less than $10^{-16}$ to $m = 160$,
giving the upper bound in the above series.

\begin{figure}
  \includegraphics[width=\columnwidth]{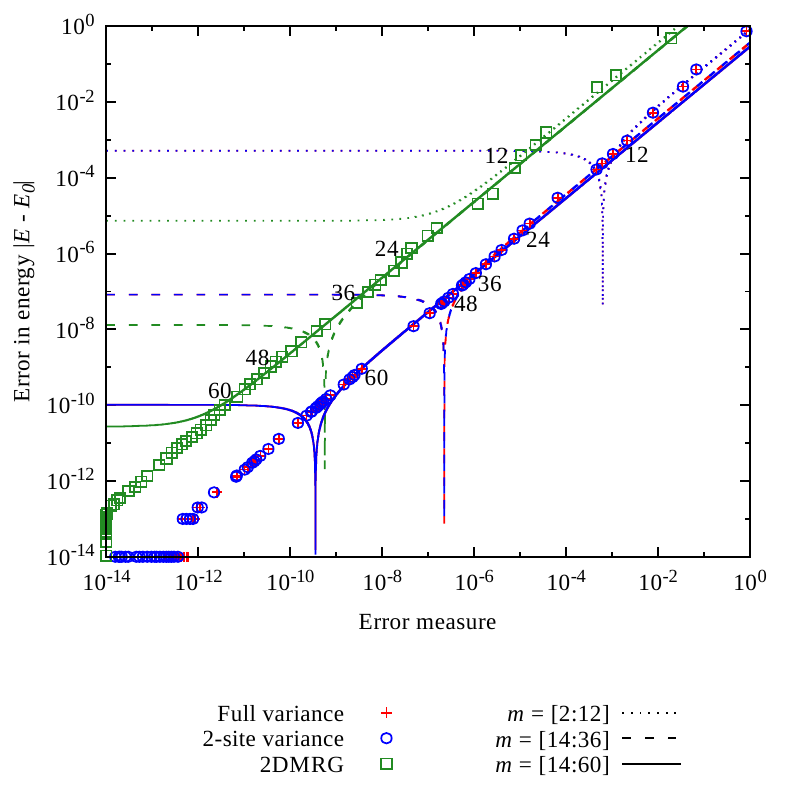}
  \caption{\label{fig:examples:heisenberg}(Color online) Energy
    differences from the true ground state and error measures
    (cf.~Sec.~\ref{sec:examples:intro}) for the $L=30$ Heisenberg
    chain. Both the 2DMRG truncation error and the error measure based
    on either the full or 2-site variance result as expected in
    essentially straight lines of slope $1$ in the log-log plot. The
    full and 2-site variance overlap completely due to the short-range
    Hamiltonian.}
\end{figure}

In Fig.~\ref{fig:examples:heisenberg} we plot the energy differences
to the ground state over the three error measures. Three linear
extrapolations in the ranges $m \in [2, 12]$, $m \in [14, 36]$ and
$m \in [14, 60]$ are done. The first range represents the case of only
a bad, low-precision calculation being available, the other ranges
showcase the increased precision attainable and -- as usually done --
exclude the lowest-precision data points.

The first extrapolation provides a much improved extrapolated
ground-state energy estimate with 2DMRG as compared to the two methods
based on 1DMRG. This is likely due to the increased effective bond
dimension in 2DMRG. In later extrapolations, all extrapolated
ground-state energies are within an order of magnitude. 

To summarize the Heisenberg model, we find mostly identical behavior
between an extrapolation in the variance and and the 2DMRG truncation
error. The latter sometimes provides better data, likely due to the
larger effective bond dimension. Generally, extrapolation can lower
the energy difference $|E-E_0|$ from the true ground state by an order
of magnitude.

\subsection{\label{sec:examples:hubbard}Hubbard chain}

\begin{figure}
  \includegraphics[width=\columnwidth]{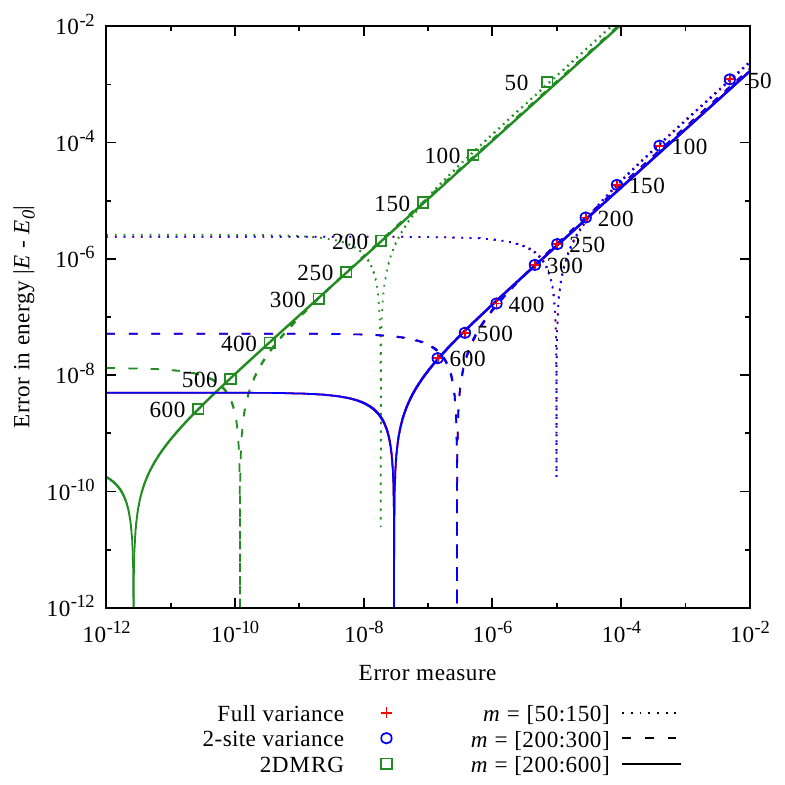}
  \caption{\label{fig:examples:hubbard}(Color online) Energy
    differences from the true ground state and error measures
    (cf.~Sec.~\ref{sec:examples:intro}) for the $L=50$ Hubbard
    chain. Three extrapolation regions were selected and the three
    error measures mostly lead to equally valid extrapolations towards
    zero error.  The full and 2-site variance overlap completely due
    to the short-range Hamiltonian. The error in the energy is reduced
    by approximately an order of magnitude via the extrapolation.}
\end{figure}

\begin{figure}[t]
  \includegraphics[width=\columnwidth]{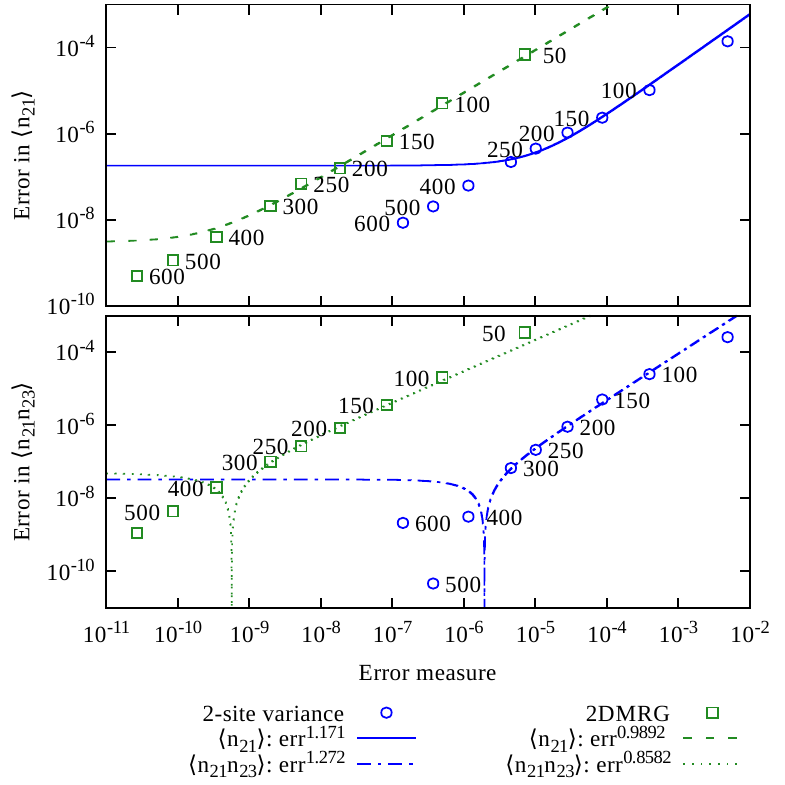}
  \caption{\label{fig:examples:hubbard:n}(Color online) Exemplary
    behavior of $\langle \hat n_{21} \rangle$ (top) and
    $\langle\hat n_{21} \hat n_{23}\rangle$ (bottom) in the Hubbard
    chain. Fits of the form $a\cdot\mathrm{err}^b + c$ in the interval
    $m \in [100, 300]$ are shown with the exponent $b$ ranging from
    $0.85$ to $1.27$ in these cases. Since the 2-site variance
    correctly calculates the full variance in this nearest-neighbor
    case, it is not shown in the figure.}
\end{figure}

As a second system, we consider the Fermi-Hubbard model on a chain of
$L=50$ sites with open boundary conditions. The Hubbard-$U$ parameter
is set to $8$, $t = 1$:
\begin{align}
  \hat H = & -\sum_{i=1}^{49} \left( \hat c^\dagger_i \cdot \hat c_{i+1} + \mathrm{h.c.} \right) + \frac{8}{2} \sum_{i=1}^{50} \left( \hat n_i^2 - \hat n_i \right) \quad.
\end{align}
Both the $\mathrm{U}(1)_N$ and $\mathrm{SU}(2)_S$ symmetries are
implemented (leading to two-component spinors $\hat c$ and
$\hat c^\dagger$). We select the sector $N=40, S=0$ for the
ground-state search. The reference value for the ground-state energy
$E_0 = 30.4096693772556 \pm 3 \times 10^{-13}$ is provided by a
calculation at $m=2000$, the test calculations are run at
$m = 50, 100, 150, 200, 250, 300, 400, 500$ and $600$ .

Extrapolations over the regions $m \in [50,150]$, $m \in [200,300]$
and $m \in [200,600]$ show that at small accuracies, all
extrapolations have roughly equal errors. At higher accuracies, the
2DMRG again benefits from its larger effective bond dimension. As
expected, the full variance and two-site variance approximation
coincide again, as this Hamiltonian also only has nearest-neighbor
interactions. All extrapolations lead to equally valid results and
lower the error in energy by approximately an order of magnitude.

In addition to the error in energy, we also consider the observables
$\langle \hat n_{21} \hat n_{23}\rangle$ and
$\langle \hat n_{21} \rangle$, again compared to a reference value
evaluated at $m = 2000$. The following noteworthy observations can be
made: first, $\langle \hat n_{21} \hat n_{23}\rangle$ seems to be
accurate down to approx.~$10^{-9}$ with the error saturating
there. Second, in the extrapolation of this observable, both 2DMRG and
the 2-site variance only lower the measured error by approximately a
factor of two. Third, when evaluating $\langle \hat n_{21} \rangle$,
the 2DMRG results are consistently more accurate than those from
1DMRG, possibly due to the two-site optimization applying better to
this observable. Fourth, the optimal coefficients $b$ in a fit of the
form $a\cdot\mathrm{err}^b + c$ range from $0.85$ to $1.27$ in these
examples. Overall, while extrapolation of these observables is more
difficult, it is still possible to obtain improved estimates for the
true value at the ground state.

\subsection{\label{sec:examples:longrange}Long-range Coulomb interactions}

\begin{figure}[t]
  \includegraphics[width=\columnwidth]{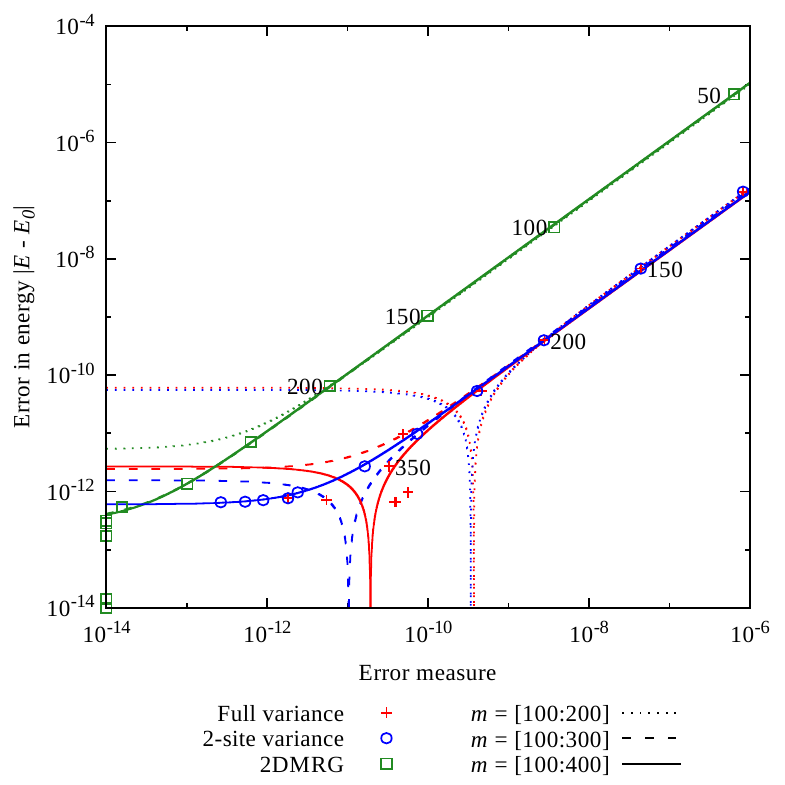}
  \caption{\label{fig:examples_longrange}(Color online) Energy
    differences from the true ground state and error measures
    (cf.~Sec.~\ref{sec:examples:intro}) for the model with long-range
    Couloumb-like interactions. While the full variance and two-site
    approximation thereof differ, both lead to comparable
    extrapolations which in turn are comparable to those based on
    2DMRG.}
\end{figure}

As an example of long-range Hamiltonians which will lead to a
difference between the full variance and the two-site approximation of
the variance, we selected an electronic model with Coulomb-like
long-range interactions. The Hamiltonian, again implementing both the
$\mathrm{U}(1)_N$ and $\mathrm{SU}(2)_S$ symmetries, is given by
\begin{equation}
  \hat H = \sum_{i=1}^{19} \hat c_i^\dagger \cdot c_{i+1} + \mathrm{h.c.} + \sum_{i=1}^{20} \sum_{j=i}^{20} \frac{1}{1 + |i-j|} \hat n_i \hat n_j \quad.
\end{equation}
$N=30$ electrons with total spin $S=0$ were placed in the system. Due
to the strong repulsion and relatively small system size, solutions
exhaust numerical accuracy around $m \approx 400$ with the reference
value $E_0 = 111.43149155591837\pm 5 \times 10^{-12}$ evaluated at
$m = 600$. Test calculations are run at
$m = 50, 100, 150, \ldots, 500$ and extrapolations are for
$m \in [100, 200]$, $m \in [100, 300]$ and $m \in [100, 400]$.

Apart from a minor advantage enjoyed by 2DMRG due to the momentarily
larger bond dimensions, the results largely coincide between 2DMRG,
the full variance and the 2-site variance. In particular, the
extrapolations based on the full variance and the two-site variance
mostly coincide very well and lower the error in the energy again by
approximately an order of magnitude.

\subsection{\label{sec:examples:hubbard-2d}Hubbard cylinder}

\begin{figure}[t]
  \includegraphics[width=\columnwidth]{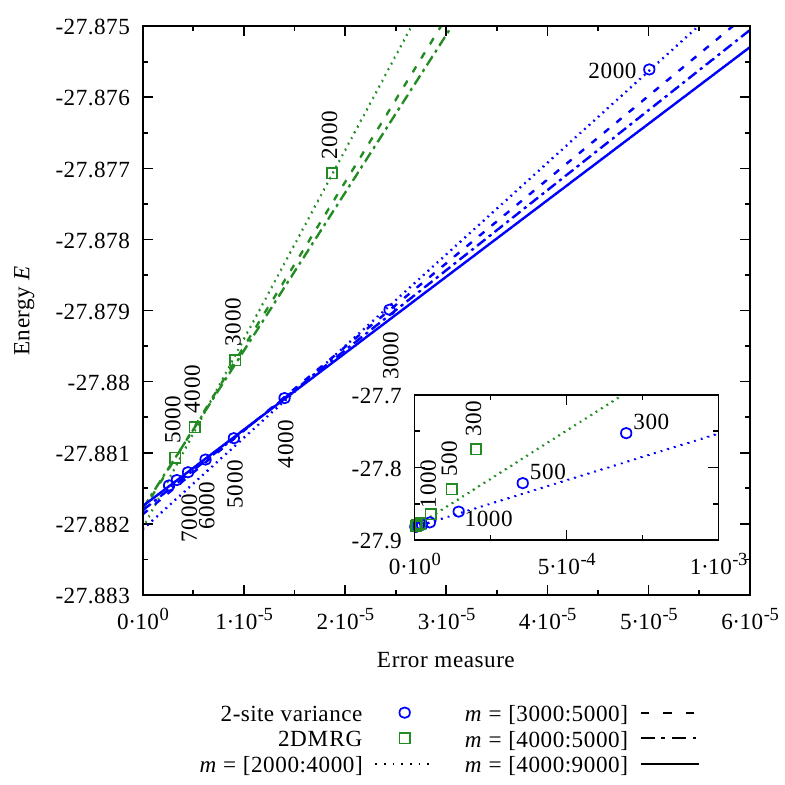}
  \caption{\label{fig:examples:hubbard-2d}(Color online) Extrapolation
    of energies towards zero error (cf.~Sec.~\ref{sec:examples:intro})
    in the Hubbard model on a cylinder. Main plot: Values used for
    extrapolation at $m \geq 2000$. Inset: Measured energies and
    errors at all bond dimensions with the first extrapolation over
    $m \in [2000,4000]$ for comparison.}
\end{figure}

\begin{figure}[t]
  \includegraphics[width=\columnwidth]{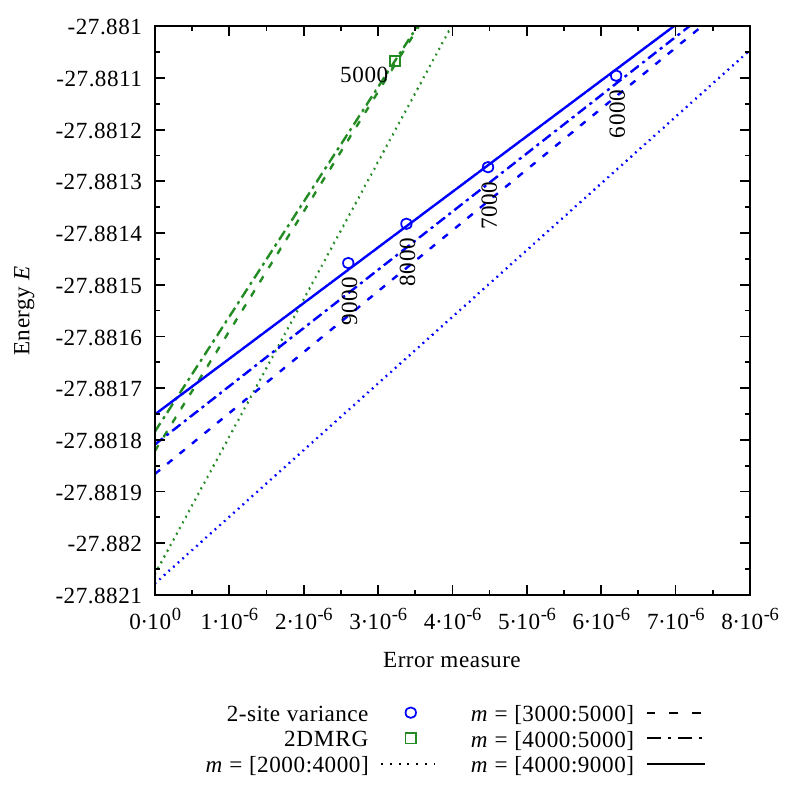}
  \caption{\label{fig:examples:hubbard-2d-zoom}(Color online) Zoomed
    version of Fig.~\ref{fig:examples:hubbard-2d}. The consecutive
    corrections of the ground-state energy estimates based on more and
    more data are clearly visible. Again, extrapolated energies {\em
      increase} with bond dimension since extrapolated energies, as
    opposed to calculated energies, do not obey a variational
    principle.}
\end{figure}

As the final example, we will attempt to calculate the ground-state
energy of the Hubbard model on a cylinder of size $10 \times
4$. $U(1)_N$ particle number conservation, $\mathrm{SU}(2)_S$ total
spin symmetry and $\mathbb{Z}_{4, k}$ quasi-momentum conservation on
the cylinder were implemented.  The Hubbard-$U$ parameter is set to
$8$, $t = 1$:
\begin{align}
  \hat H = & -\sum_{x=1}^{10} \sum_{\alpha=1}^4 2 \cos\left( 2 \pi \frac{\alpha}{4} \right)  \hat c^\dagger_{x, \alpha} \cdot \hat c_{x, \alpha} \nonumber \\
         & -\sum_{x=1}^{9} \sum_{\alpha=1}^4 \left( \hat c^\dagger_{x, \alpha} \cdot \hat c_{x+1,\alpha} + \mathrm{h.c.} \right) \nonumber \\
         & + \frac{8}{2} \sum_{x=1}^{10} \sum_{\alpha=1}^4 \left[ \sum_{\beta \gamma = 1}^4 \frac{1}{4} \hat c^\dagger_{x, \alpha} \cdot \hat c_{x, \beta} \times \hat c^\dagger_{x, \gamma} \cdot \hat c_{x, \alpha - \beta + \gamma} \right] \nonumber \\
  & - \frac{8}{2} \sum_{x=1}^{10} \sum_{\alpha=1}^4 \hat c^\dagger_{x, \alpha} \cdot \hat c_{x, \alpha} \quad.
\end{align}
$\hat c^{(\dagger)}_{x,\alpha}$ is the two-component spinor
annihilating (creating) an electron on ring $x$ with momentum
$\alpha$. The two last lines implement the real-space on-site
interaction which, in momentum space along each ring, becomes
long-ranged. $N=36$ electrons with total spin $S=0$ and momentum $k=0$
were placed in the system.

\begin{table}[b!]
  \caption{\label{tab:hubbard-2d}Resulting energy expectation values
    at $m=5000$ and $m=9000$ and extrapolated ground-state energy
    estimates for the Hubbard model on a cylinder.}
  \begin{ruledtabular}
    \begin{tabular}{l|cc}
      $m$            & 1DMRG \& 2-site var. & 2DMRG \\
      \hline \hline
      $5000$         & $-27.8807953$        & $-27.8810672$ \\
      $9000$         & $-27.8814580$        & --            \\
      $[2000, 4000]$ & $-27.8820779$        & $-27.8820598$ \\
      $[3000, 5000]$ & $-27.8818661$        & $-27.8818215$ \\
      $[4000, 5000]$ & $-27.8818090$        & $-27.8817841$ \\
      $[4000, 9000]$ & $-27.8817508$        & --
    \end{tabular}
\end{ruledtabular}
\end{table}

Due to the complicated system, no exact reference calculation was
possible. 2DMRG was run at bond dimension
$m = 300, 500, 1000, 2000, 3000, 4000$ and $5000$. With 1DMRG,
increasing $m$ further to $m = 6000, 7000, 8000$ and $m = 9000$ was
possible due to lower computational effort and less memory usage. The
$m=9000$ $\mathrm{SU}(2)$-invariant states correspond to roughly
$24'000$ states if only symmetries without inner multiplicity were
used. Evaluating the full variance was not possible due to the large MPO
bond dimension.

In the extrapolations, the initial data at $m < 1000$ was discarded
and extrapolations over $m \in [2000, 4000]$, $m \in [3000, 5000]$ and
$m \in [4000, 5000]$ were done for both 2DMRG and 1DMRG data as well
as $m \in [4000, 9000]$ for 1DMRG data. Both extrapolations in 2DMRG
truncation error and the two-site variance initially underestimated
the ground-state energy with later extrapolations correcting the
estimate slightly upwards. Extrapolations at $m \in [2000, 4000]$ and
$m \in [3000, 5000]$ give nearly the same results between the 2-site
variance and 2DMRG.

The 2-site variance further corrects the estimate upwards for
$m \in [4000, 5000]$ and $m \in [4000, 9000]$ resulting in a
highest-precision estimate for the ground-state energy of
$-27.8817508$ compared to the highest-precision 2DMRG estimate of
$-27.8817841$ (cf.~Tab.~\ref{tab:hubbard-2d}).

\section{\label{sec:conclusions}Conclusions}

We have shown that the variance itself as well as its two-site
approximation allow for a reliable extrapolation of observables to the
ground state from a series of small-$m$ states. Measuring the two-site
approximation of the variance is considerably cheaper than evaluating
the full variance and leads to valid extrapolations comparable in
quality to those resulting from 2DMRG. It hence allows the use of
1DMRG with its significant speed-up and reduced memory usage over the
traditional 2DMRG.

All extrapolations encountered here lower the error in energy by
approximately one order of magnitude from the most precise data point
available, consistent with previous
observations.\cite{white05:_densit} We must stress, however, that the
variational property of DMRG is lost if we use \emph{any} sort of
extrapolation.

\begin{acknowledgments} We would like to thank T.~K\"ohler,
  S.~Paeckel, L.~Schoonderwoerd, E.~M.~Stoudenmire and S.~R.~White for
  very helpful discussions.  C.~H.~acknowledges funding through the
  ExQM graduate school and the Nanosystems Initiative Munich. J.~H.~is
  supported by the European Commission via ERC grant nr. 715861
  (ERQUAF).
\end{acknowledgments}


%

\end{document}